\documentclass[aps,prl,twocolumn,superscriptaddress, floatfix,nofootinbib]{revtex4-2}  \usepackage{graphicx}  \usepackage{dcolumn}   \usepackage{bm}        \usepackage{amssymb}   \usepackage[caption=false]{subfig}

\newcommand{\phref}[2]{\href{#2}{#1}}
\begin{document}

\title{High-statistics measurement of neutrino quasielastic-like scattering at 
$\sim 6~GeV$ 
on a hydrocarbon target }
\newcommand{\Rutgers}{Rutgers, The State University of New Jersey, Piscataway, New Jersey 08854, USA}
\newcommand{\Hampton}{Hampton University, Dept. of Physics, Hampton, VA 23668, USA}
\newcommand{\Dortmund}{Institute of Physics, Dortmund University, 44221, Germany }
\newcommand{\Otterbein}{Department of Physics, Otterbein University, 1 South Grove Street, Westerville, OH, 43081 USA}
\newcommand{\JMU}{James Madison University, Harrisonburg, Virginia 22807, USA}
\newcommand{\Florida}{University of Florida, Department of Physics, Gainesville, FL 32611}
\newcommand{\UCIrvine}{Department of Physics and Astronomy, University of California, Irvine, Irvine, California 92697-4575, USA}
\newcommand{\CBPF}{Centro Brasileiro de Pesquisas F\'{i}sicas, Rua Dr. Xavier Sigaud 150, Urca, Rio de Janeiro, Rio de Janeiro, 22290-180, Brazil}
\newcommand{\PUCP}{Secci\'{o}n F\'{i}sica, Departamento de Ciencias, Pontificia Universidad Cat\'{o}lica del Per\'{u}, Apartado 1761, Lima, Per\'{u}}
\newcommand{\INRM}{Institute for Nuclear Research of the Russian Academy of Sciences, 117312 Moscow, Russia}
\newcommand{\Jlab}{Jefferson Lab, 12000 Jefferson Avenue, Newport News, VA 23606, USA}
\newcommand{\Pittsburgh}{Department of Physics and Astronomy, University of Pittsburgh, Pittsburgh, Pennsylvania 15260, USA}
\newcommand{\Guanajuato}{Campus Le\'{o}n y Campus Guanajuato, Universidad de Guanajuato, Lascurain de Retana No. 5, Colonia Centro, Guanajuato 36000, Guanajuato M\'{e}xico.}
\newcommand{\Athens}{Department of Physics, University of Athens, GR-15771 Athens, Greece}
\newcommand{\Tufts}{Physics Department, Tufts University, Medford, Massachusetts 02155, USA}
\newcommand{\WM}{Department of Physics, College of William \& Mary, Williamsburg, Virginia 23187, USA}
\newcommand{\FNAL}{Fermi National Accelerator Laboratory, Batavia, Illinois 60510, USA}
\newcommand{\Purdue}{Department of Chemistry and Physics, Purdue University Calumet, Hammond, Indiana 46323, USA}
\newcommand{\MCLA}{Massachusetts College of Liberal Arts, 375 Church Street, North Adams, MA 01247}
\newcommand{\UMD}{Department of Physics, University of Minnesota -- Duluth, Duluth, Minnesota 55812, USA}
\newcommand{\Northwestern}{Northwestern University, Evanston, Illinois 60208}
\newcommand{\UNI}{Universidad Nacional de Ingenier\'{i}a, Apartado 31139, Lima, Per\'{u}}
\newcommand{\Rochester}{University of Rochester, Rochester, New York 14627 USA}
\newcommand{\Austin}{Department of Physics, University of Texas, 1 University Station, Austin, Texas 78712, USA}
\newcommand{\USM}{Departamento de F\'{i}sica, Universidad T\'{e}cnica Federico Santa Mar\'{i}a, Avenida Espa\~{n}a 1680 Casilla 110-V, Valpara\'{i}so, Chile}
\newcommand{\Geneva}{University of Geneva, 1211 Geneva 4, Switzerland}
\newcommand{\Chicago}{Enrico Fermi Institute, University of Chicago, Chicago, IL 60637 USA}
\newcommand{\hired}{}
\newcommand{\OregonState}{Department of Physics, Oregon State University, Corvallis, Oregon 97331, USA}
\newcommand{\oxford}{Oxford University, Department of Physics, Oxford, OX1 3PJ United Kingdom}
\newcommand{\umiss}{University of Mississippi, Oxford, Mississippi 38677, USA}
\newcommand{\upenn}{Department of Physics and Astronomy, University of Pennsylvania, Philadelphia, PA 19104}
\newcommand{\AMU}{AMU Campus, Aligarh, Uttar Pradesh 202001, India}
\newcommand{\wroclaw}{University of Wroclaw, plac Uniwersytecki 1, 50-137 Wrocław, Poland}
\newcommand{\Mohali}{Department of Physical Sciences, IISER Mohali, Knowledge City, SAS Nagar, Mohali - 140306, Punjab, India}
\newcommand{\CINVESTAV}{Departamento de Fisica Col. San Pedro Zacatenco, 07360 Mexico, DF, Av. Instituto Politécnico Nacional, Mexico}
\newcommand{\york}{Department of Physics and Astronomy, Toronto, Ontario, M3J 1P3 Canada}
\newcommand{\mateusfcarneiroThanks}{Now at Brookhaven National Laboratory}

\author{M.F.~Carneiro}\thanks{\mateusfcarneiroThanks}  \affiliation{\OregonState} \affiliation{\CBPF}
\author{D.~Ruterbories}                   \affiliation{\Rochester}
\author{Z.~Ahmad~Dar}                 \affiliation{\AMU}
\author{F.~Akbar}                         \affiliation{\AMU}
\author{D.A.~Andrade}                     \affiliation{\Guanajuato}
\author{M.~V.~Ascencio}                   \affiliation{\PUCP}
\author{W.~Badgett}
\affiliation{\FNAL}
\author{A.~Bashyal}                       \affiliation{\OregonState}
\author{A.~Bercellie}                     \affiliation{\Rochester}
\author{M.~Betancourt}                    \affiliation{\FNAL}
\author{K.~Bonin}                         \affiliation{\UMD}
\author{A.~Bravar}                        \affiliation{\Geneva}
\author{H.~Budd}                          \affiliation{\Rochester}
\author{G.~Caceres}                       \affiliation{\CBPF}
\author{T.~Cai}                           \affiliation{\Rochester}
\author{H.~da~Motta}                      \affiliation{\CBPF}
\author{G.A.~D\'{i}az~}                   \affiliation{\Rochester}  \affiliation{\PUCP}
\author{J.~Felix}                         \affiliation{\Guanajuato}
\author{L.~Fields}                        \affiliation{\FNAL}
\author{A.~Filkins}                       \affiliation{\WM}
\author{R.~Fine}                          \affiliation{\Rochester}
\author{A.M.~Gago}                        \affiliation{\PUCP}
\author{A.~Ghosh}                         \affiliation{\USM}  \affiliation{\CBPF}
\author{R.~Gran}                          \affiliation{\UMD}
\author{D.~Hahn}                          \affiliation{\FNAL}
\author{D.A.~Harris}                      \affiliation{\york}  \affiliation{\FNAL}
\author{S.~Henry}                         \affiliation{\Rochester}
\author{J.~Hylen}
\affiliation{\FNAL}
\author{S.~Jena}                          \affiliation{\Mohali}
\author{D.~Jena}                           \affiliation{\FNAL}
\author{C.~Joe}  \affiliation{\FNAL}
\author{B.~King} \affiliation{\FNAL}
\author{J.~Kleykamp}                      \affiliation{\Rochester}
\author{M.~Kordosky}                      \affiliation{\WM}
\author{D.~Last}                          \affiliation{\upenn}
\author{T.~Le}                            \affiliation{\Tufts}  \affiliation{\Rutgers}
\author{J.~LeClerc}                       \affiliation{\Florida}
\author{A.~Lozano}                        \affiliation{\CBPF}
\author{X.-G.~Lu}                         \affiliation{\oxford}
\author{E.~Maher}                         \affiliation{\MCLA}
\author{S.~Manly}                         \affiliation{\Rochester}
\author{W.A.~Mann}                        \affiliation{\Tufts}
\author{K.S.~McFarland}                   \affiliation{\Rochester}
\author{C.L.~McGivern}  \affiliation{\FNAL}  \affiliation{\Pittsburgh}
\author{A.M.~McGowan}                     \affiliation{\Rochester}
\author{B.~Messerly}                      \affiliation{\Pittsburgh}
\author{J.~Miller}                        \affiliation{\USM}
\author{J.G.~Morf\'{i}n}                  \affiliation{\FNAL}
\author{M.~Murphy}
\affiliation{\FNAL}
\author{D.~Naples}                        \affiliation{\Pittsburgh}
\author{J.K.~Nelson}                      \affiliation{\WM}
\author{C.~Nguyen}                       \affiliation{\Florida}
\author{A.~Norrick}                       \affiliation{\WM}
\author{A.~Olivier}                       \affiliation{\Rochester}
\author{V.~Paolone}                       \affiliation{\Pittsburgh}
\author{G.N.~Perdue}                      \affiliation{\FNAL}  \affiliation{\Rochester}
\author{P.~Riehecky}
\affiliation{\FNAL}
\author{H.~Schellman}                     \affiliation{\OregonState}
\author{P.~Schlabach}
\affiliation{\FNAL}
\author{C.J.~Solano~Salinas}              \affiliation{\UNI}
\author{H.~Su}                            \affiliation{\Pittsburgh}
\author{M.~Sultana}                       \affiliation{\Rochester}
\author{V.S.~Syrotenko}                   \affiliation{\Tufts}
\author{D.~Torretta}                      \affiliation{\FNAL}
\author{C.~Wret}                          \affiliation{\Rochester}
\author{B.~Yaeggy}                        \affiliation{\USM}
\author{K.~Yonehara}
\affiliation{\FNAL}
\author{L.~Zazueta}                       \affiliation{\WM}

\date{\today}

\newcommand{\minerva}{MINERvA}

\begin{abstract}
We measure neutrino charged current quasielastic-like scattering on hydrocarbon at high statistics using the wide-band NuMI beam with neutrino energy peaked at 6 GeV.  The double-differential cross section is reported in terms of muon longitudinal ($p_\parallel$) and transverse ($p_\perp$) momentum.   Cross-section contours versus lepton momentum components are approximately described by a conventional generator-based simulation, however discrepancies are observed for transverse momenta above 0.5 GeV/c for longitudinal momentum ranges 3 to 5 GeV/c and 9 to 20 GeV/c.   The single differential cross section versus momentum transfer squared ($d\sigma/dQ_{QE}^2$) is measured over a four-decade range of $Q^2$ that extends to $10~GeV^2$.  The cross section turn-over and fall-off in the $Q^2$ range 0.3 to $10~GeV^2$ is not fully reproduced by generator predictions that rely on dipole form factors.  Our measurement probes the axial-vector content of the hadronic current and complements the electromagnetic form factor data obtained using electron-nucleon elastic scattering. 
These results help oscillation experiments because they probe the importance of various correlations and final-state interaction effects within the nucleus, which have different effects on the visible energy in detectors.
\end{abstract}

\pacs{FIX THIS }  
\maketitle

The Charged Current Quasi-Elastic (CCQE) neutrino interaction ({\em i.e.} $\nu_{\mu}n\rightarrow\mu^{-}p$) is an important channel 
in the $E_\nu$ range of a few GeV and is of value in searches for leptonic CP-symmetry violation \cite{Abe:2018wpn, Abe:2017vif, Acero:2019ksn, NOvA:2018gge, Acciarri:2015uup, Abe:2015zbg}.  
Because there is little missing energy, this channel allows a good estimate of the incident neutrino energy. 
However,  imperfect knowledge of nuclear effects remains a limiting factor for oscillation measurements \cite{Alvarez-Ruso:2017oui}. These uncertainties are significant in current experiments
\cite{Abe:2018wpn, Abe:2017vif, Acero:2019ksn, NOvA:2018gge} and will become more important with the statistics of 
DUNE~\cite{Acciarri:2015uup} and Hyper-Kamiokande~\cite{Abe:2015zbg}.

For free nucleons, quasielastic scattering is described by the standard theory of weak interactions combined with nucleon form factors \cite{Llewellyn:1972}. Electron-nucleon scattering experiments \cite{Bradford:2006yz} measure the electromagnetic form factors, but measurement of the axial-vector form factor, F$_A$,  at four-momentum transfer squared $Q^2 \sim 0.1$ GeV$^2$ can only be done via $\nu/\bar{\nu}$ nucleon scattering.   

The axial-vector form factor is usually parameterized using the dipole form and has been measured at zero energy transfer through beta-decay experiments \cite{Wilkinson:1982, Maerkisch:2014nma}. The vector (V), axial-vector (A), and VA interference terms of free-nucleon hadronic currents
have been studied on free or quasi-free nucleons on hydrogen and deuterium targets
\cite{Miller:1982,Kitagaki:1990,Kitagaki:1983px,Allasia:1990uy}. 

Neutrino oscillation experiments in the few-GeV range, however, use detectors constructed of 
carbon \cite{Acero:2019ksn, AguilarArevalo:2008qa}, oxygen \cite{Abe:2011ks}, iron \cite{Adamson:2014pgc} or argon \cite{Acciarri:2015uup,Adams:2018sgn}. Nuclear effects are significant and must be modeled for these experiments to reach their full physics potential. Historically, a Relativistic Fermi Gas (RFG) \cite{Smith:1972} has been used to model the intial state nucleon but modifications are necessary to reproduce experimental data \cite{AguilarArevalo:2008qa, Fiorentini:2013ezn, Abe:2014iza, Acero:2019ksn}.  The Local Fermi Gas (LFG) is an extension to the RFG with a local density approximation \cite{Negele:1970,Maruhn:2009}. Alternatively, Spectral Functions (SF) techniques \cite{Cenni:1997} use a mean field to replace the sum of individual interactions.

 Long-range correlations between nucleons are modeled using a Random Phase Approximation (RPA) correction~\cite{Nieves:2004wx,Martini:2009,Graczyk:2003ru,Singh:1992, Martini:2016eec,Nieves:2017lij} 
  to account for the screening effect that arises from the proximity of other nucleons in the nuclear potential well.
 The RPA correction reduces the interaction rate at low $Q^2$ while enhancing moderate $Q^2$ interactions.

A wide range of two-particle, two-hole models using a meson-exchange formalism are tested against electron scattering (e,e') data \cite{Benhar:2006er, Benhar:2006wy, Zeller:1973ge, Barreau:1983ht, OConnell:1987kww, Bagdasaryan:1988hp, Sealock:1989nx}.  Attempts to predict the neutrino rate and {\it pp} and {\it pn} knockout rate are given in~\cite{Martini:2009,Nieves:2011,Megias:2016fjk,VanCuyck:2017wfn}.  This analysis uses a simulation with the Valencia 2p2h model \cite{Nieves:2011}.

A complete description of the experimental signature for quasielastic scattering 
must also account for the propagation through the nucleus of particles produced by any initial charged-current interaction. The charged lepton produced escapes the nucleus without interacting but final state hadrons are likely to interact.   Such final-state interactions (FSI) may produce new particles such as pions or mimic the CCQE signal through absorption of pions in resonance production. In both cases the observed final state differs from the original interaction. 

We use a topology-based signal definition where a muon, zero or more nucleons, and no mesons or heavy baryons are in the final state (CCQE-like). 
CCQE-like processes include pion production where the pion is absorbed in the nucleus and 2p2h processes where more than one nucleon is produced. The history of CCQE measurements is extensive \cite{Gran:2006jn,Lyubushkin:2008pe,Aguilar-Arevalo:2013dva,Fields:2013zhk,Fiorentini:2013ezn,Walton:2014esl,Wolcott:2015hda,Betancourt:2017uso,Acciarri:2014gev,Abe:2014iza,Abe:2015oar,Abe:2016tmq,Patrick:2018gvi,Abe:2018pwo,Ruterbories:2018gub}, but the community has yet to converge on a full description of the nuclear effects since the measured final state is determined by a mixture initial interaction dynamics and nuclear effects.

In this Letter we report a study of muon neutrino CCQE-like interactions 
in the NuMI \cite{Adamson:2015dkw} "Medium Energy" (ME) beam.  The data correspond to an exposure of $1.061\times10^{21}$ protons on target (POT), which combined with the higher flux per POT results in over a factor of ten increase in statistics above our previous measurements~\cite{Fields:2013zhk, Fiorentini:2013ezn,Patrick:2018gvi,Ruterbories:2018gub}. The new configuration provides a broad neutrino flux peaked at $6$ GeV. 
We present two-dimensional cross sections for CCQE-like scattering as a function of muon transverse ($p_\perp$) and longitudinal ($p_\parallel$) momentum. We also report the differential cross section versus the square of the momentum transferred using a quasielastic interaction hypothesis, 
where $Q^2_{QE} = 2E_\nu(E_\mu-p_\parallel)-M_\mu^2$ 
and the neutrino energy $E_\nu$ is also determined using the QE hypothesis (see \cite{Ruterbories:2018gub}). 
This result extends the $Q^2_{QE}$ range by a factor of four compared to previous 
measurements. 
The NuMI beam line consists of a $120$-GeV primary proton beam, a two-interaction-length graphite target, two parabolic focusing horns, and a $675$-m decay pipe. For these data, taken between 2013-2017, 
the horn polarities are set to create a neutrino-dominated beam.
The beam line is modeled with a Geant4-based~\cite{Agostinelli:2002hh,Allison:2006ve}  simulation (g4numi~\cite{Aliaga:2016oaz} version 6, built against Geant version v.9.4.p2). There are known discrepancies between Geant4 predictions of proton-on-carbon and other interactions relevant to NuMI flux predictions. MINERvA 
corrects the Geant4 flux predictions with hadron-production data \cite{Aliaga:2016oaz}. In addition, measurements of neutrino-electron ($\nu-e$) scatters, as described in \cite{Valencia:2019mkf},
constrain the flux and reduces the normalization uncertainty on the integrated flux between 2 and 20 GeV from $7.8\%$ 
to $3.9\%$. 

We restrict this study to events originating in the central scintillator tracker region of the \minerva~ detector~\cite{Aliaga:2013uqz}.  The target mass consists of $88.5\%$, $8.2\%$, and $2.5\%$ carbon, hydrogen, and oxygen, respectively, plus small amounts of heavier nuclei.  The 5.3-tonne tracker fiducial region is followed by an electromagnetic calorimeter made up of 20 scintillator planes interleaved with 0.2-cm thick lead sheet, followed by a hadronic calorimeter region of 20 scintillator planes interleaved with 2.54-cm thick iron slabs. 
The magnetized MINOS muon spectrometer \cite{Michael:2008bc} begins 2 m downstream and provides momentum and charge information for muons.

Neutrino interactions are simulated using the GENIE 2.12.6 event generator \cite{Andreopoulos:2009rq}. The GENIE default interaction model is adjusted to match MINERvA GENIE tune v1 (MnvGENIEv1). This model includes three modifications to the default GENIE model. First, the Valencia RPA correction \cite{Nieves:2004wx,Gran:2017psn}, appropriate for a Fermi gas \cite{Martini:2009,Nieves:2017lij}, is added as a function of energy and three momentum transfer. 
Second, the prediction for multi-nucleon scattering given by the Valencia model \cite{Nieves:2011pp,Gran:2013kda,Schwehr:2016pvn}  in GENIE 2.12.6 is added and modified with an empirical fit~\cite{Rodrigues:2015hik} based on previous MINERvA data. The modification, referred to as the "low recoil fit," increases the integrated 2p2h rate by 49\%. Finally, non-resonant pion production is reduced by $57\%$ to agree with a fit to measurements of that process on deuterium~\cite{Rodrigues:2016xjj}.

The kinematics of each interaction are reconstructed using the measured muon momentum and angle with respect to the beam
as described in \cite{Ruterbories:2018gub}. To address the MINOS acceptance 
only events with muons created within  $<20^{o}$ of the neutrino beam and above 1.5 GeV/c in momentum are accepted. 

As a cross-check of its flux predictions, MINERvA also uses samples of neutrino-nucleus interactions with less than $800$~MeV transferred to the hadronic system. Data and simulation comparisons show a discrepancy as a function of neutrino energy.  To determine the source of this discrepancy we fit the neutrino energy distributions in different spatial regions of the detector to templates that allow both the beamline parameters (i.e. focusing horn current and position) and the muon energy scale to float.  Hadron production and neutrino interaction uncertainties are evaluated to obtain the systematic uncertainty on the fit results.  The data/simulation prediction before and after the muon energy scale shift are shown in the supplement.  
The discrepancy is most consistent with a 3.6\% muon energy scale shift, which is 1.8 times the {\em a priori} energy scale uncertainty.  In this analysis, the reconstructed muon energy is shifted by 3.6\%, with an uncertainty of 1.0\% (the posterior uncertainty from the fit).  

We retain two populations of events: a muon-only sample with no identified proton and a muon+proton sample. These samples are analyzed separately since their background components have different sources. For both of these  
populations,  there are three sidebands used to constrain three backgrounds, as described in \cite{Ruterbories:2018gub}.

As the signal definition for CCQE-like  includes no final state mesons or heavy baryons, the energy loss profiles of tracks contained within MINERvA are required to be consistent with a proton hypothesis.  For events with $Q^{2}_{QE}> 0.6$~GeV$^{2}$ the proton-interaction probability is high, so no energy-loss cut is made in this region.  This results in a small discontinuity in the transverse momentum distributions for the muon+additional track samples.  
To reduce inelastic backgrounds, 
events with untracked energy above $0.5$ GeV are removed.  Events with Michel electron (from the decay chain $\pi^{\pm}\rightarrow\mu^{\pm}\rightarrow e^{\pm}$) are also vetoed.

\begin{figure}[h]
    \centering
\includegraphics[trim={0.4cm 0.6cm 2.5cm 0.5cm},clip,height=3.5cm]{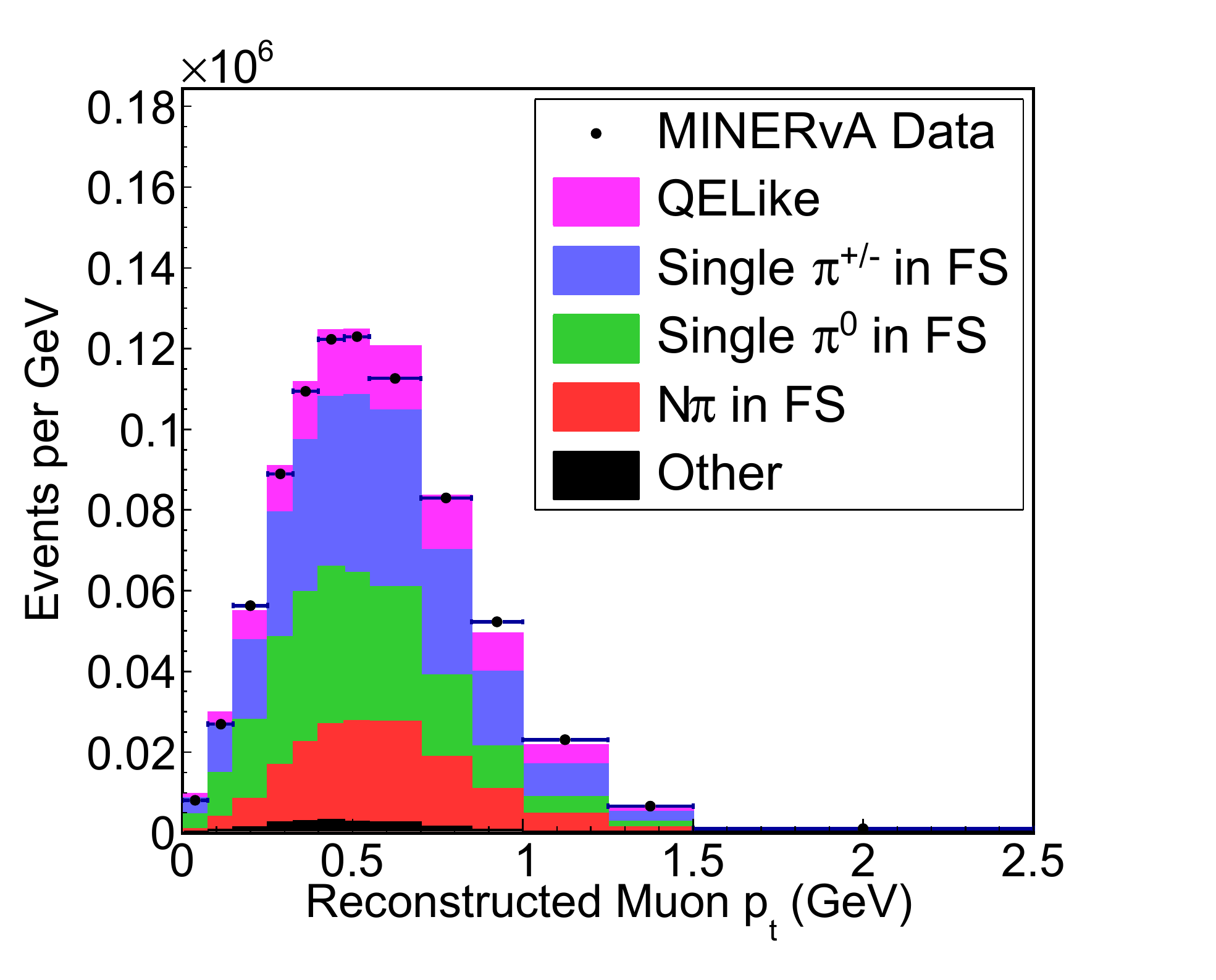}
\includegraphics[trim={1.3cm 0.6cm 2.5cm 0.5cm},clip,height=3.5cm]{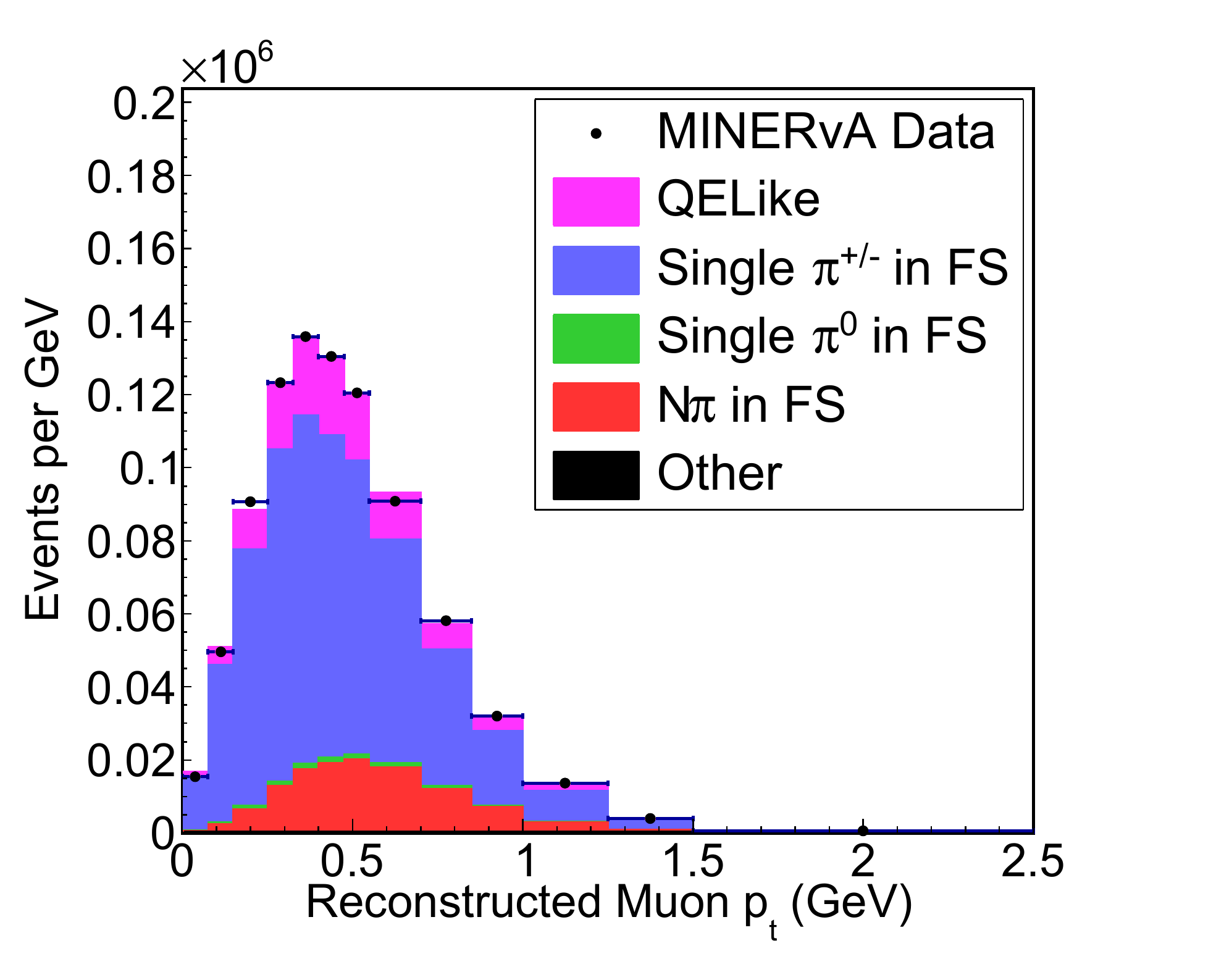}
    \caption{1-track sideband $p_T$ distributions for data and predictions after fitting, for (left) $\pi^0$ and (right) $\pi^\pm$ Michel candidates.}
    \label{fig:sidebands}
\end{figure}
\begin{figure}[]

    \includegraphics[trim={0cm 0.6cm 2.5cm 0.7cm},clip,height=3.6cm]{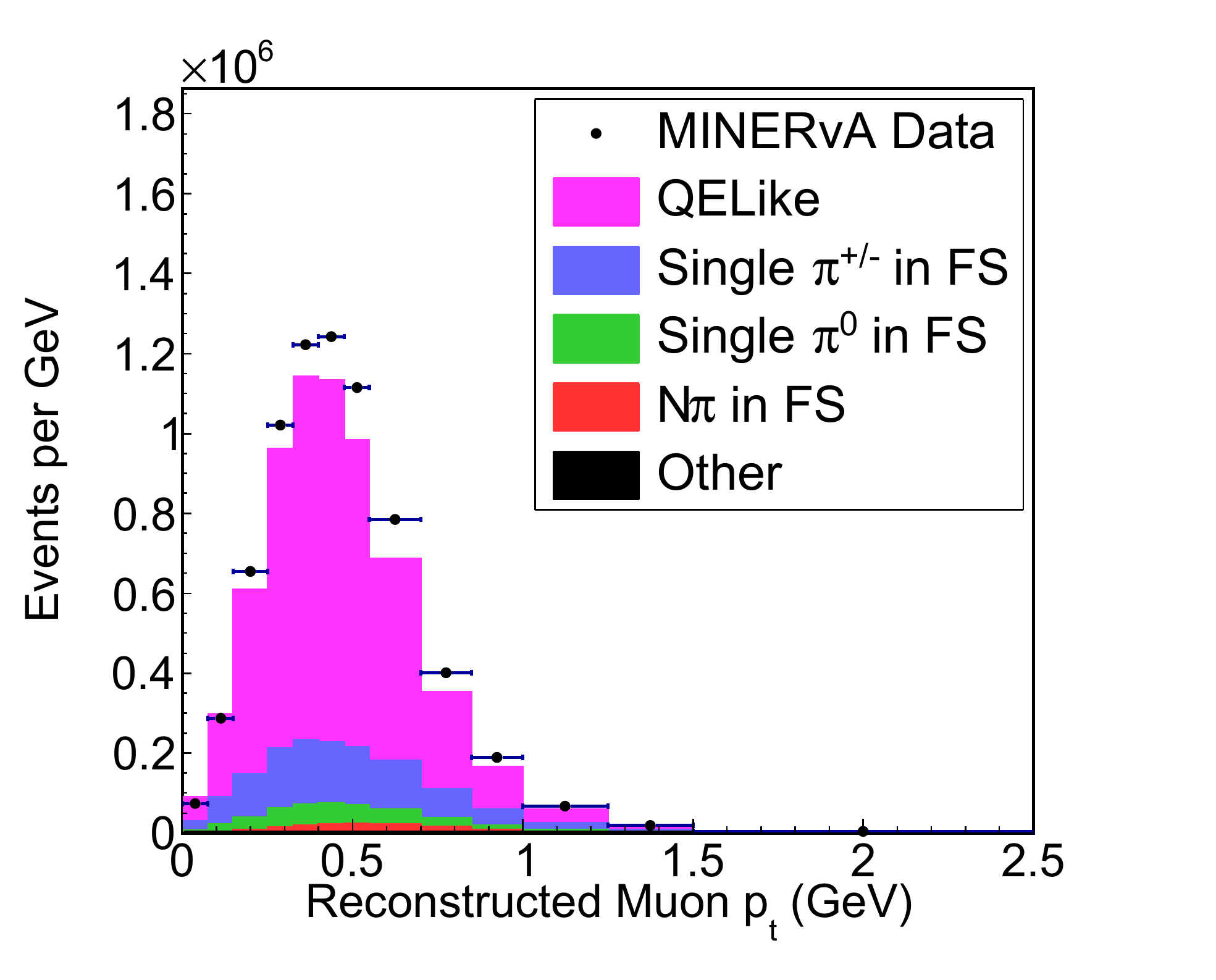}
    \includegraphics[trim={1.8cm 0.6cm 2.5cm 0.7cm},clip,height=3.6cm]{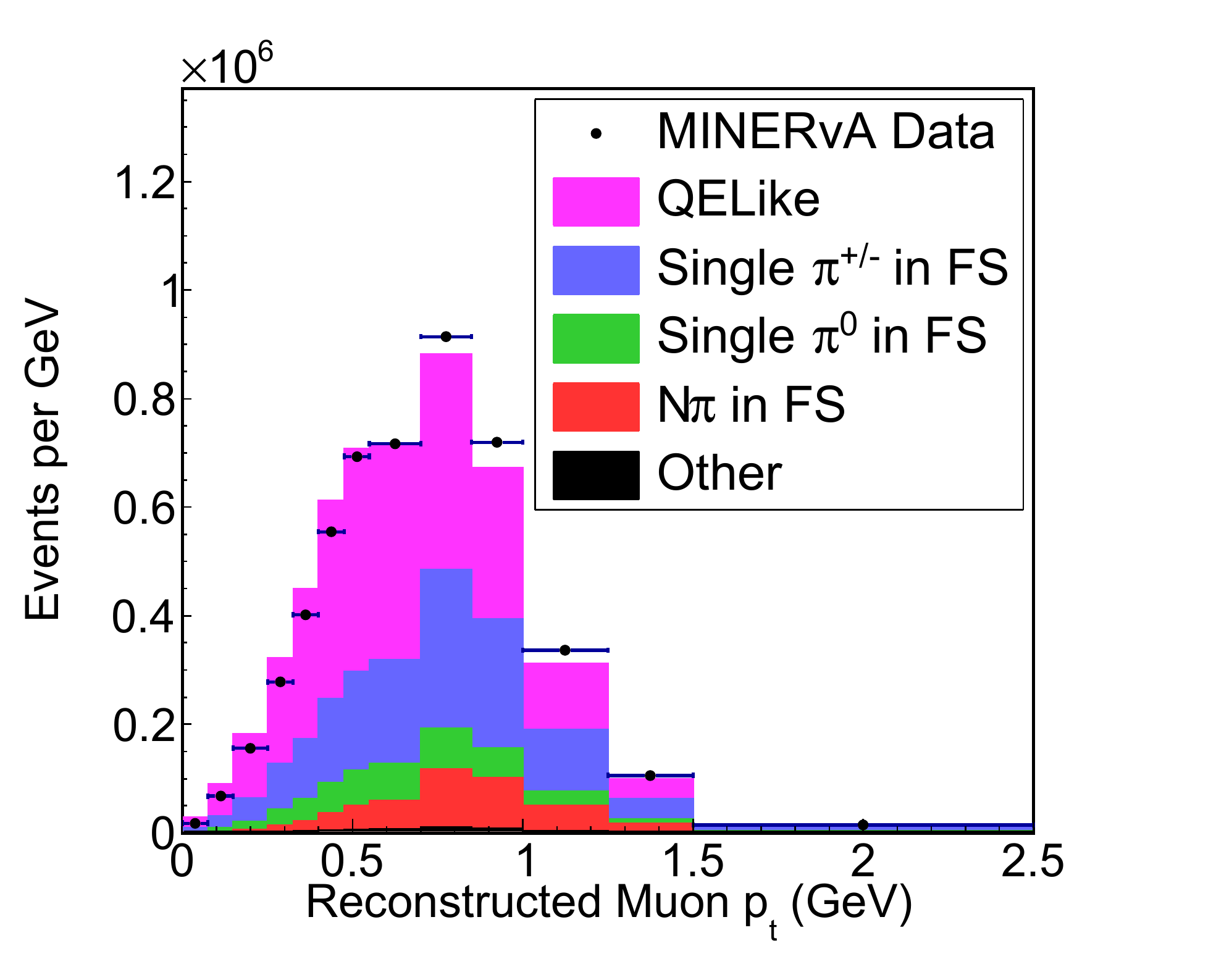}
       \caption{Reconstructed muon transverse momentum in (left) 1-track and (right) 2+track  signal samples. The primary background in both samples comes from charged current pion production. \label{sigonly_after_fit}}
\end{figure}

The first sideband consists of events having two or more clusters of energy detached from the primary vertex but passing all other cuts.   This sample, shown in Fig.~\ref{fig:sidebands} (left), helps constrain backgrounds from processes with $\pi^{0}$s in the final state or events where a $\pi^{+}$ charge exchanges.
The second sideband consists of events passing all cuts but the Michel electron cut.  
This sample is primarily sensitive to backgrounds from charged pions, as shown in Fig.~\ref{fig:sidebands} (right). 
The third (and smallest) sideband comes from events with both a Michel electron and extra clusters, and it is sensitive to multi-pion events.

To constrain the background predictions, simultaneous fits are made to the three sidebands as a function of muon transverse momentum, for the single-track and the multi-track topologies separately.   Templates based on three simulated background distributions are  fit to the data and the resulting three background normalizations for each topology are used to estimate the contamination from each source.  The effect of the fit on the backgrounds versus muon transverse momentum is shown in the supplement.    Using the fit results we subtract the predicted backgrounds from the data in each bin.   The one- and multi-track signal samples have $670,022$ and $648,518$ events, respectively, and are shown in Fig.~\ref{sigonly_after_fit} with the predicted backgrounds after the fit.

After background subtraction the data are unfolded, following the method of  D'Agostini  \cite{DAgostini:1994fjx,DAgostini:2010hil}, via the implementation in RooUnfold \cite{Adye:2011gm} using $4$ iterations. To minimize model dependence, the unfolded $Q^2_{QE}$ is the one calculated with the true muon kinematics assuming a quasielastic hypothesis, not the generator-level momentum transfer squared.   
The unfolded sample is corrected for selection efficiency as predicted by the simulation. The selection has an average efficiency of $70\%$ in bins inside the edges of the phase space. The efficiency is  approximately $70\%$ below 0.1 GeV$^2$ in  $Q^{2}_{QE}$, reducing to $10\%$ at 10 GeV$^2$. The efficiency-corrected distributions are normalized by the integral of the predicted neutrino flux in the $0 - 120$ GeV range and by the number of nucleons 
($3.23\times 10^{30}$ in the fiducial region) to derive differential cross sections.

The cross section uncertainties for four representative $p_\parallel$ bins are shown in Fig.~\ref{Sys2Dsmall}.  Uncertainties for remaining bins and for the $Q^2_{QE}$ result are available in the supplement.  Muon reconstruction uncertainties, which include muon energy scale, resolution and angle uncertainties, dominate in most bins. 
A description of the remaining uncertainty classes and how they are assessed can be found in \cite{Ruterbories:2018gub}.  Additionally, we add an uncertainty to account for the possibility of low-$Q^{2}$ suppression in pion events, evaluated by adding the low-$Q^{2}$ suppression described in \cite{Stowell:2019zsh} to our default model. The flux uncertainties are described in \cite{Valencia:2019mkf}.  
\begin{figure}
             
             \includegraphics[width=\linewidth]{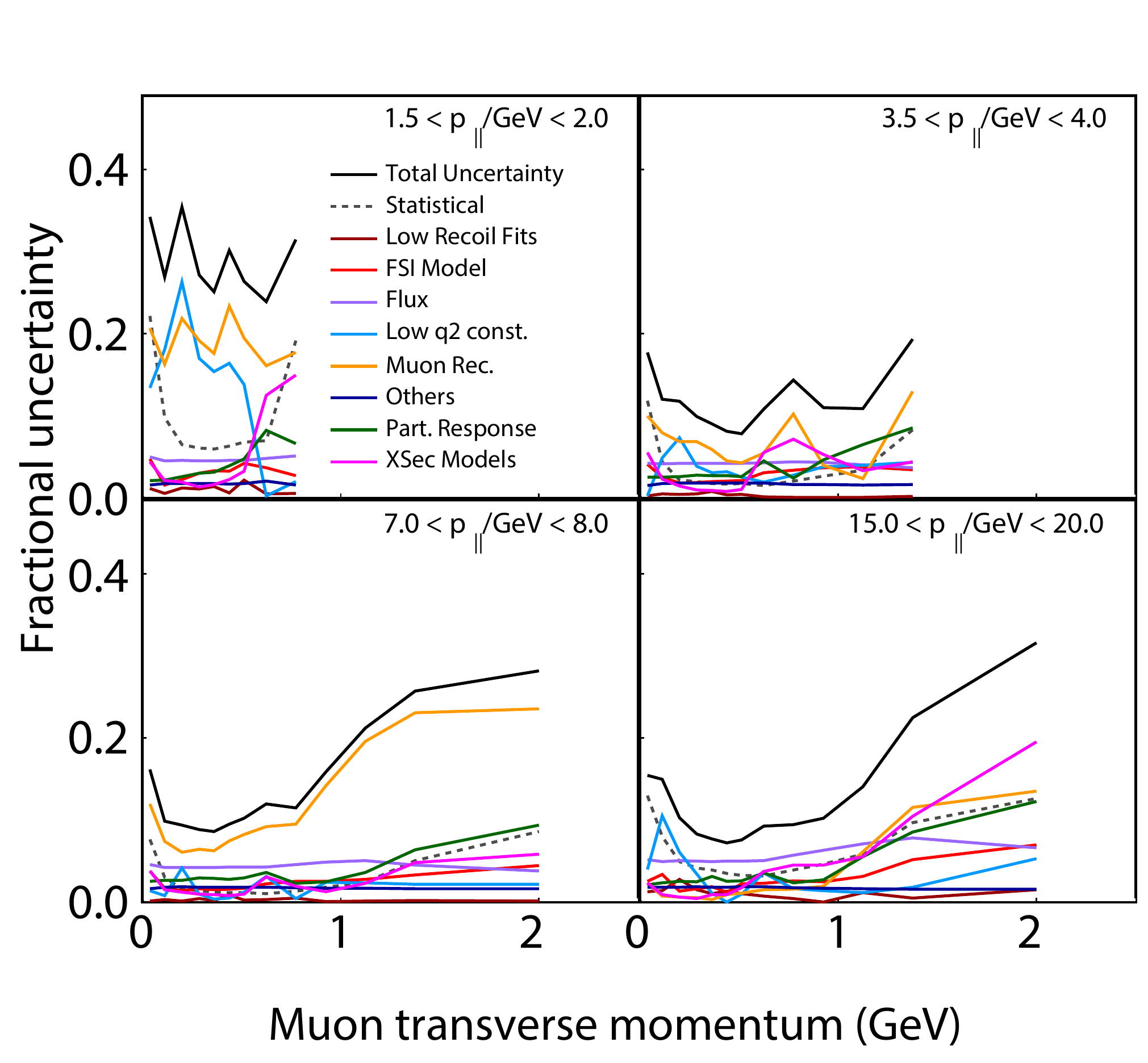}
             
    \caption{Fractional systematic uncertainty on the double-differential cross section as a function of $p_\perp$ and $p_\parallel$.}
     \label{Sys2Dsmall} 
\end{figure}
\begin{figure*}
    \centering
    \includegraphics[width=\linewidth]{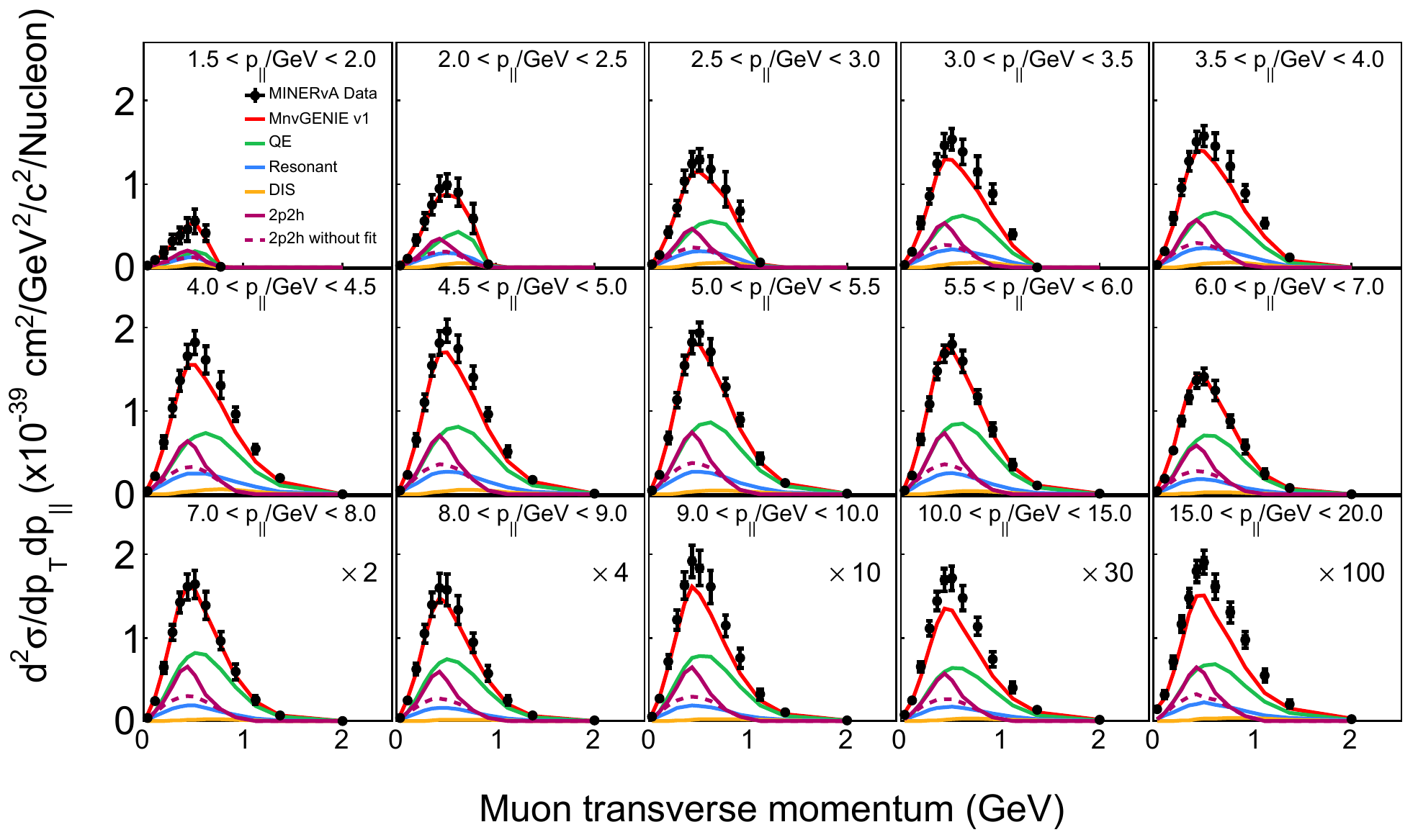}
    \caption{$d^2\sigma/dp_\perp/dp_\parallel$ for data and the MnvGENIEv1 reference simulation in bins of $p_\parallel$.  The predictions for the contributions to the final-state signal channel from CCQE, resonant, DIS and 2p2h processes are also shown.}
    \label{XSec2D}
\end{figure*}

The double-differential cross section is presented in Fig.~\ref{XSec2D}.
Here, MnvGENIEv1 serves as a reference simulation to which the data are compared.	The simulation is seen to reproduce the data at zeroth order, but discrepancies are apparent.  Bins above the spectral peak in $p_{T}$ are underpredicted in the $p_{\parallel}$ range 3.0 to 5.0 GeV.   From 5.5 to 8.0 GeV the distributions below the spectral peak are overpredicted; underprediction of event rate resumes dramatically at $p_\parallel$ above 9.0 GeV.  The simulation shows that CCQE and 2p2h comprise the dominant spectral components, and that discrepancies could be alleviated by modest adjustments, particularly for CCQE at higher $p_T$.

\begin{figure}[h]
    \centering
\includegraphics[width=\linewidth]{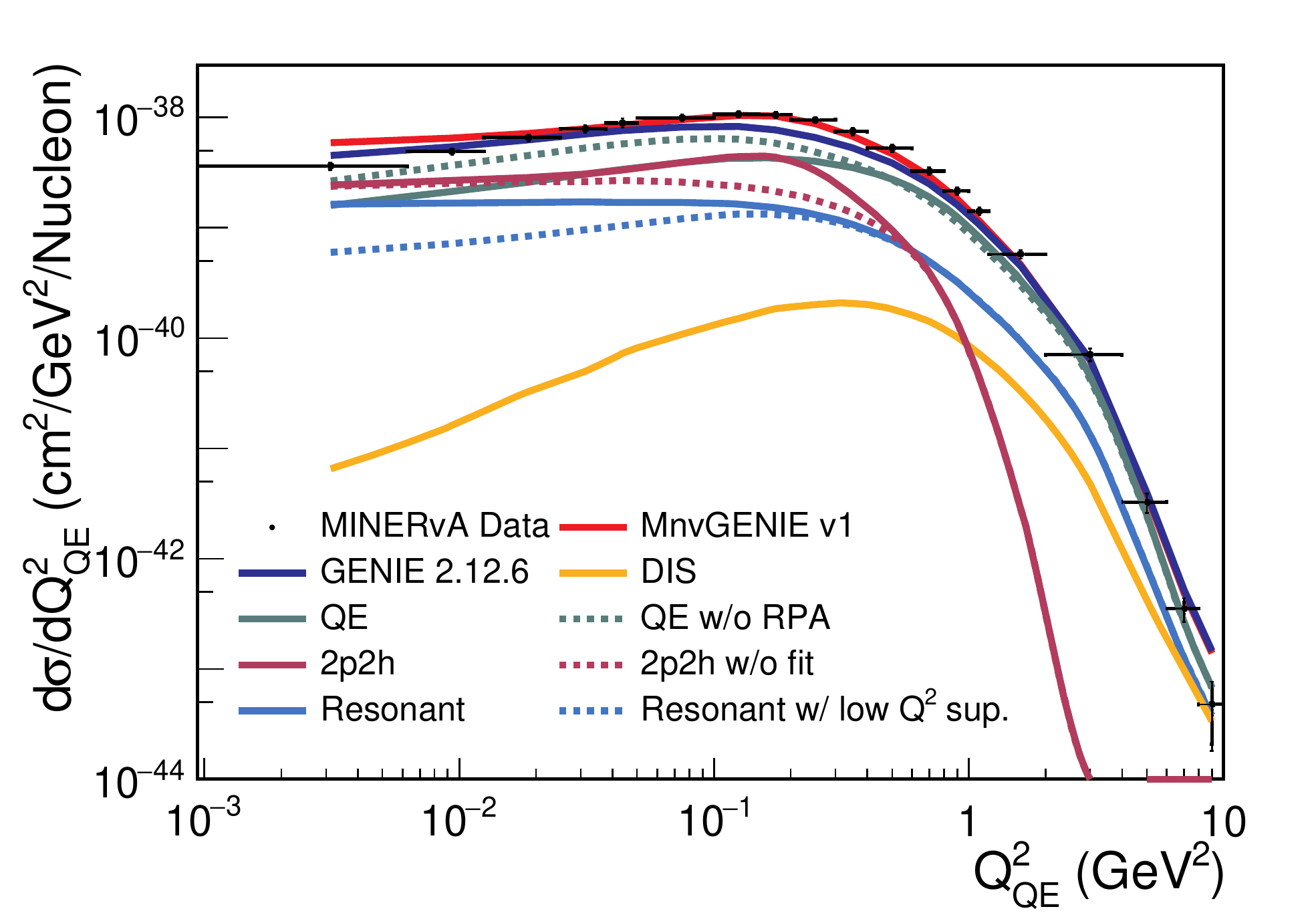}
\includegraphics[width=\linewidth]{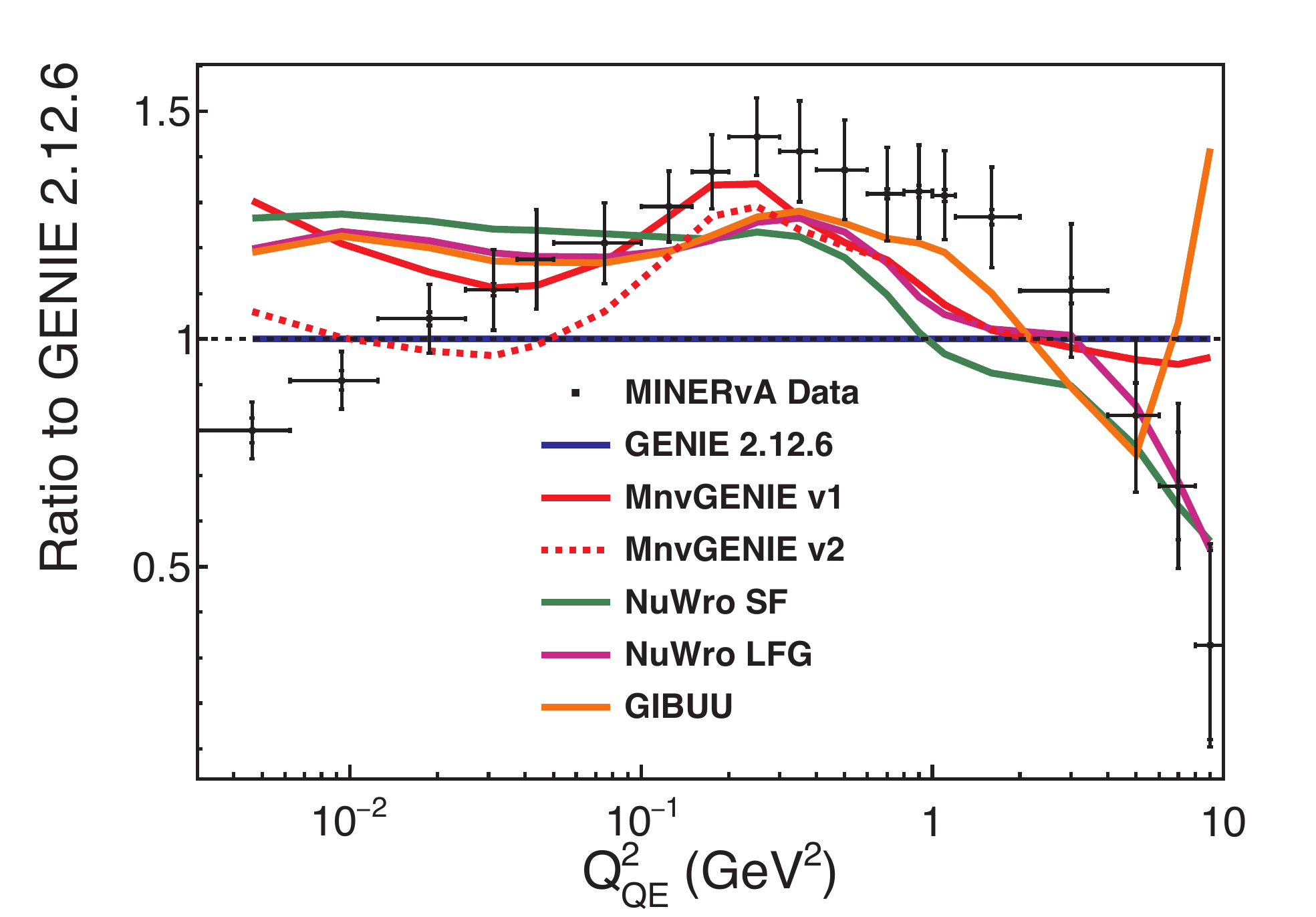}
    \caption{Top:  Differential cross section as a function of $Q^2$. Bottom:  Generator predictions compared to data. All are plotted as ratio to the predictions of unmodified GENIE 2.12.6.
        }
    \label{XSec1D}
\end{figure}

The single-differential cross section $d\sigma/dQ_{QE}^2$ is presented in Fig.~\ref{XSec1D} (top).    
The fall-off of the cross section for $Q^2 > 1.0$ GeV$^2$ is reproduced at moderate and high $Q^2$ by the MnvGENIEv1 reference simulation, indicating that dipole forms for the 
vector and axial vector nucleon form factors remain appropriate.
Fig.~\ref{XSec1D} (bottom) shows the ratio of data and selected generators to the reference simulation.   Here the cross-section turnover in the range 0.3 to $\sim 3.0$ GeV$^2$ proceeds more gradually than
predicted; all generators under-predict the data throughout this region.   These general features
are similar to those observed for the electromagnetic form factors in electron-nucleon elastic scattering experiments (see Fig. 17 of~\cite{Puckett:2011xg}).   The present work, by mapping neutrino quasielastic scattering into the 
multi-GeV $Q^2$ region, provides new information about the axial-vector part of the nucleon current that cannot be accessed by electron scattering.   This new information will enable tests of nuclear models heretofore based solely on electron scattering~\cite{Meyer:2016oeg,Qattan:2015qxa}.

\begin{table}
\renewcommand*{\arraystretch}{1.2}
\begin{ruledtabular}
\begin{tabular}{lcc}
Model &	 $\chi^2$ - linear &	 $\chi^2$ - log  \\ 
\hline 
GENIE 2.12.6 &	        1031 &	       1543 \\
\qquad+RPA+$\pi$ tune &	        420 &	        927 \\
\qquad+RPA+$\pi$ tune &   &  \\
\qquad\qquad +MINOS $\pi$ low $Q^2_{QE}$ sup. &	        403 &	        986 \\
\hline
GENIE 2.12.6 + 2p2h & 2299 &  1913 \\
\qquad+RPA+$\pi$ tune + recoil fit &	 &	 \\
\qquad\qquad (MnvGENIEv1) & 1194 & 1155 \\
\qquad+RPA+$\pi$ tune &	       1068 &	       1221 \\
\qquad+RPA+$\pi$ tune+ recoil fit  & & \\
\qquad\qquad + MINOS $\pi$ low $Q^2_{QE}$ sup. &  870 & 989 \\
\qquad + recoil fit +$\pi$ tune &	       2714 &	       2052 \\
\qquad + RPA +$\pi$ tune + recoil fit &  &  \\
 \qquad\qquad + MINERvA $\pi$ low $Q^2_{QE}$ sup. &  &  \\
\qquad\qquad  (MnvGENIEv2) &	        799 &	        953 \\
\hline
NuWro SF &	       3533 &	       6188 \\
NuWro LFG &	       3176 &	       5914 \\
\hline
GiBUU &	       1729 &	       1890 \\
\end{tabular}
\end{ruledtabular}
\caption{\label{2DChi2} $\chi^2$ of model variants compared to $d^2\sigma\over dp_\perp dp_\parallel$. Both standard and log-normal $\chi^2$ are shown; the number of degrees of freedom for each comparison is 184.}
\end{table}

Tab.~\ref{2DChi2} provides the $\chi^2$ for model predictions of the $p_\perp-p_\parallel$ differential cross section measurement. 
The models differ in additional effects added to the default version of the GENIE generator. The variations denoted ``+RPA" include the Valencia RPA model \cite{Nieves:2004wx,Gran:2017psn}, while ``+2p2h" adds the Valencia prediction for the multi-nucleon  scattering~\cite{Nieves:2011pp,Gran:2013kda,Schwehr:2016pvn}. ``+MINOS (\minerva) $\pi$ low $Q^{2}_{QE}$ sup." refers to an empirical resonant pion low-$Q^2_{QE}$ suppression based on MINOS~\cite{Adamson:2014pgc} (\minerva~\cite{Stowell:2019zsh}) data.  ``$\pi$ tune" refers to a 57\% reduction non-resonant pion production motivated by deuterium data~\cite{Rodrigues:2016xjj}.  

In general, the $\chi^2$ values for all of the models are poor, but the models with the smallest $\chi^2$ are those that include RPA but not 2p2h.  This is in contrast with previous \minerva~measurements~\cite{Ruterbories:2018gub} of this channel in the lower-energy NuMI tune, indicating that the expanded phase space of this dataset is illuminating regions of mismodeling that could not be seen in prior measurements.  A $\chi^2$ table for model predictions of the single-differential cross section versus $Q^{2}_{QE}$ is available in the Supplement.

This result is the first CCQE-like measurement at $Q^2_{QE}$ above 4~GeV$^2$ and spans almost four orders of magnitude in $Q^2$.    The data in this high-$Q^2$ region diverge from most predictions that are based on generators used by current oscillation experiments, and there are no models that are even in approximate agreement over all ranges of $Q^2$.  
The high-statistics double-differential cross sections 
will be an important benchmark for model developers who tune models for future neutrino oscillation measurements.  

\begin{acknowledgments}
This document was prepared by members of the MINERvA Collaboration using the resources of the Fermi National Accelerator Laboratory (Fermilab), a U.S. Department of Energy, Office of Science, HEP User Facility. Fermilab is managed by Fermi Research Alliance, LLC (FRA), acting under Contract No. DE-AC02-07CH11359. These resources included support for the \minerva~ construction project, and support for construction also was granted by the United States National Science Foundation under Award No. PHY-0619727 and by the University of Rochester. Support for participating scientists was provided by NSF and DOE (USA); by CAPES and CNPq (Brazil); by CoNaCyT (Mexico); by Proyecto Basal FB 0821, CONICYT PIA ACT1413, Fondecyt 3170845 and 11130133 (Chile); by CONCYTEC, DGI-PUCP, and IDI/IGI-UNI (Peru); and by the Latin American Center for Physics (CLAF).  We thank the MINOS Collaboration for use of its near detector data. Finally, we thank the staff of Fermilab for support of the beam line, the detector, and computing infrastructure.
\end{acknowledgments}


\begin{thebibliography}{99}

\bibitem{Abe:2018wpn} 
  K.~Abe {\it et al.} [T2K Collaboration],
    Phys.\ Rev.\ Lett.\  {\bf 121}, no. 17, 171802 (2018)
  [arXiv:1807.07891 [hep-ex]].
      
\bibitem{Abe:2017vif} 
  K.~Abe {\it et al.} [T2K Collaboration],
    Phys.\ Rev.\ D {\bf 96}, no. 9, 092006 (2017)
  Erratum: [Phys.\ Rev.\ D {\bf 98}, no. 1, 019902 (2018)]
    PhysRevD.98.019902
  [arXiv:1707.01048 [hep-ex]].
          
        \bibitem{Acero:2019ksn} 
  M.~A.~Acero {\it et al.} [NOvA Collaboration],
    Phys.\ Rev.\ Lett.\  {\bf 123}, no. 15, 151803 (2019)
   [arXiv:1906.04907 [hep-ex]].
    \bibitem{NOvA:2018gge} 
  M.~A.~Acero {\it et al.} [NOvA Collaboration],
    Phys.\ Rev.\ D {\bf 98}, 032012 (2018)
   [arXiv:1806.00096 [hep-ex]].
          
\bibitem{Acciarri:2015uup} 
  R.~Acciarri {\it et al.} [DUNE Collaboration],
    arXiv:1512.06148 [physics.ins-det].
    
\bibitem{Abe:2015zbg} 
  K.~Abe {\it et al.} [Hyper-Kamiokande Proto- Collaboration],
    \phref{PTEP {\bf 2015}, 053C02 (2015)}
    [arXiv:1502.05199 [hep-ex]].
    
        
      
\bibitem{Alvarez-Ruso:2017oui} 
  L.~Alvarez-Ruso {\it et al.},
    Prog.\ Part.\ Nucl.\ Phys.\  {\bf 100}, 1 (2018)
    [arXiv:1706.03621 [hep-ph]].
    
\bibitem{Llewellyn:1972}
  C.~H.~Llewellyn Smith, 
  Phys.\ Rept.\ {\bf 3}, 261 (1972). 
  
\bibitem{Bradford:2006yz} 
  R.~Bradford, A.~Bodek, H.~S.~Budd and J.~Arrington,
    Nucl.\ Phys.\ Proc.\ Suppl.\  {\bf 159}, 127 (2006)
    [hep-ex/0602017].
    
\bibitem{Wilkinson:1982}
  D.~H.~Wilkinson, 
  Nucl.\ Phys.\ {\bf A377}, 474 (1982).

\bibitem{Maerkisch:2014nma} 
  B.~Maerkisch and H.~Abele,
    arXiv:1410.4220 [hep-ph].
    \bibitem{Miller:1982}
  K.~L.~Miller {\it et al.},
  Phys. Rev. \b{D26}, 537 (1982).


\bibitem{Kitagaki:1990}
  T.~Kitagaki {\it et al.},
  Phys.\ Rev.\ {\bf D42}, 1331 (1990).

\bibitem{Kitagaki:1983px} 
  T.~Kitagaki {\it et al.},
    Phys.\ Rev.\ D {\bf 28}, 436 (1983).
      \bibitem{Allasia:1990uy} 
  D.~Allasia {\it et al.},
    Nucl.\ Phys.\ B {\bf 343}, 285 (1990).
      
\bibitem{AguilarArevalo:2008qa} 
  A.~A.~Aguilar-Arevalo {\it et al.} [MiniBooNE Collaboration],
    Nucl.\ Instrum.\ Meth.\ A {\bf 599}, 28 (2009)
    [arXiv:0806.4201 [hep-ex]].
    \bibitem{Abe:2011ks} 
  K.~Abe {\it et al.} [T2K Collaboration],
    Nucl.\ Instrum.\ Meth.\ A {\bf 659}, 106 (2011)
    [arXiv:1106.1238 [physics.ins-det]].
    \bibitem{Adamson:2014pgc} 
  P.~Adamson {\it et al.} [MINOS Collaboration],
    Phys.\ Rev.\ D {\bf 91}, no. 1, 012005 (2015)
    [arXiv:1410.8613 [hep-ex]].
      
\bibitem{Adams:2018sgn} 
  C.~Adams {\it et al.} [MicroBooNE Collaboration],
    arXiv:1811.02700 [hep-ex].
     
\bibitem{Smith:1972}
  R.~Smith and E.~Moniz,
  Nucl.\ Phys.\ B{\bf 43}, 605 (1972).
  
\bibitem{Fiorentini:2013ezn} 
  G.~A.~Fiorentini {\it et al.} [MINERvA Collaboration],
    Phys.\ Rev.\ Lett.\  {\bf 111}, 022502 (2013)
    [arXiv:1305.2243 [hep-ex]].
     \bibitem{Abe:2014iza} 
  K.~Abe {\it et al.} [T2K Collaboration],
    Phys.\ Rev.\ D {\bf 92}, no. 11, 112003 (2015)
    [arXiv:1411.6264 [hep-ex]].
     




\bibitem{Negele:1970} 
   J.~W.~Negele,
   Phys.\ Rev.\ {\bf C1}, 1260 (1970).
  
  \bibitem{Maruhn:2009}
  J.~A.~Maruhn, P.~-G.~Reinhard, and E.~Suraud,
  {\it Simple Models of Many-Fermion Systems}
  (Springer-Verlag Berlin Heidelberg, 2009).
  
\bibitem{Cenni:1997} 
   R.~Cenni, T.~W.~Donnelly, and A.~Molinari,
   Phys.\ Rev.\ C{\bf 56}, 276 (1997).

 
\bibitem{Nieves:2004wx} 
  J.~Nieves, J.~E.~Amaro and M.~Valverde,
    Phys.\ Rev.\ C {\bf 70}, 055503 (2004)
  Erratum: [Phys.\ Rev.\ C {\bf 72}, 019902 (2005)]
    [nucl-th/0408005].
    
\bibitem{Martini:2009}
  M.~Martini, M.~Ericson, G.~Chanfray, and J.~Marteau,
  Phys.\ Rev.\ C{\bf 80}, 065501 (2009).

\bibitem{Graczyk:2003ru} 
  K.~M.~Graczyk and J.~T.~Sobczyk,
    Eur.\ Phys.\ J.\ C {\bf 31}, 177 (2003)
    [nucl-th/0303054].
    
\bibitem{Singh:1992}
  S.~K.~Singh and E.~Oset,
  Nucl.\ Phys.\ {\bf A542}, 587 (1992).

\bibitem{Martini:2016eec} 
  M.~Martini, N.~Jachowicz, M.~Ericson, V.~Pandey, T.~Van Cuyck and N.~Van Dessel,
    Phys.\ Rev.\ C {\bf 94}, no. 1, 015501 (2016)
    [arXiv:1602.00230 [nucl-th]].
    
\bibitem{Nieves:2017lij} 
  J.~Nieves and J.~E.~Sobczyk,
    Annals Phys.\  {\bf 383}, 455 (2017)
    [arXiv:1701.03628 [nucl-th]].
    
\bibitem{Egiyan:2006} 
   K.~Egiyan {\it et al.} [CLAS Collaboration],
   Phys.\ Rev.\ Lett.\ {\bf 96}, 082501 (2006).
   
\bibitem{Shneor:2007} 
   R.~Shneor {\it et al.} [Jeferson Lab Hall A  Collaboration],
   Phys.\ Rev.\ Lett.\ {\bf 99}, 072501 (2007).
   
\bibitem{Subedi:2008} 
   R.~Subedi {\it et al.},
   Science {\bf 320}, 1476 (2008).
   
\bibitem{Bodek:1981} 
   A.~Bodek and J.~L.~Ritchie,
   Phys.\ Rev.\ D{\bf 23}, 1070 (1981).


\bibitem{Nieves:2011} 
   J.~Nieves, I.~R.~Simo, and M.~J.~V.~Vacas,
   Phys.\ Rev.\ C{\bf 83}, 045501 (2011).

\bibitem{Gonzalez-Jimenez:2014eqa} 
  R.~Gonzaléz-Jiménez, G.~D.~Megias, M.~B.~Barbaro, J.~A.~Caballero and T.~W.~Donnelly,
    Phys.\ Rev.\ C {\bf 90}, no. 3, 035501 (2014)
    [arXiv:1407.8346 [nucl-th]].
      
\bibitem{Megias:2016fjk} 
  G.~D.~Megias, J.~E.~Amaro, M.~B.~Barbaro, J.~A.~Caballero, T.~W.~Donnelly and I.~Ruiz Simo,
    Phys.\ Rev.\ D {\bf 94}, no. 9, 093004 (2016)
    [arXiv:1607.08565 [nucl-th]].
    
\bibitem{VanCuyck:2017wfn} 
  T.~Van Cuyck, N.~Jachowicz, R.~González-Jiménez, J.~Ryckebusch and N.~Van Dessel,
    Phys.\ Rev.\ C {\bf 95}, no. 5, 054611 (2017)
    [arXiv:1702.06402 [nucl-th]].
    
\bibitem{Gran:2006jn} 
  R.~Gran {\it et al.} [K2K Collaboration],
    Phys.\ Rev.\ D {\bf 74}, 052002 (2006)
  doi:10.1103/PhysRevD.74.052002
  [hep-ex/0603034].
    \bibitem{Lyubushkin:2008pe} 
  V.~Lyubushkin {\it et al.} [NOMAD Collaboration],
    Eur.\ Phys.\ J.\ C {\bf 63}, 355 (2009)
    [arXiv:0812.4543 [hep-ex]].
      
\bibitem{Aguilar-Arevalo:2013dva} 
  A.~A.~Aguilar-Arevalo {\it et al.} [MiniBooNE Collaboration],
    Phys.\ Rev.\ D {\bf 88}, no. 3, 032001 (2013)
    [arXiv:1301.7067 [hep-ex]].
      
\bibitem{Fields:2013zhk} 
  L.~Fields {\it et al.} [MINERvA Collaboration],
    Phys.\ Rev.\ Lett.\  {\bf 111}, no. 2, 022501 (2013)
    [arXiv:1305.2234 [hep-ex]].
      
\bibitem{Walton:2014esl} 
  T.~Walton {\it et al.} [MINERvA Collaboration],
    Phys.\ Rev.\ D {\bf 91}, no. 7, 071301 (2015)
    [arXiv:1409.4497 [hep-ex]].
      
\bibitem{Wolcott:2015hda} 
  J.~Wolcott {\it et al.} [MINERvA Collaboration],
    Phys.\ Rev.\ Lett.\  {\bf 116}, no. 8, 081802 (2016)
    [arXiv:1509.05729 [hep-ex]].
      
\bibitem{Betancourt:2017uso} 
  M.~Betancourt {\it et al.} [MINERvA Collaboration],
    Phys.\ Rev.\ Lett.\  {\bf 119}, no. 8, 082001 (2017)
    [arXiv:1705.03791 [hep-ex]].
      
\bibitem{Acciarri:2014gev} 
  R.~Acciarri {\it et al.} [ArgoNeuT Collaboration],
    Phys.\ Rev.\ D {\bf 90}, no. 1, 012008 (2014)
    [arXiv:1405.4261 [nucl-ex]].
      
\bibitem{Abe:2015oar} 
  K.~Abe {\it et al.} [T2K Collaboration],
    Phys.\ Rev.\ D {\bf 91}, no. 11, 112002 (2015)
    [arXiv:1503.07452 [hep-ex]].
      
\bibitem{Abe:2016tmq} 
  K.~Abe {\it et al.} [T2K Collaboration],
    Phys.\ Rev.\ D {\bf 93}, no. 11, 112012 (2016)
    [arXiv:1602.03652 [hep-ex]].
      
\bibitem{Patrick:2018gvi} 
  C.~E.~Patrick {\it et al.} [MINERvA Collaboration],
    Phys.\ Rev.\ D {\bf 97}, no. 5, 052002 (2018)
    [arXiv:1801.01197 [hep-ex]].
      
\bibitem{Abe:2018pwo} 
  K.~Abe {\it et al.} [T2K Collaboration],
    Phys.\ Rev.\ D {\bf 98}, no. 3, 032003 (2018)
    [arXiv:1802.05078 [hep-ex]].
      
\bibitem{Ruterbories:2018gub} 
  D.~Ruterbories {\it et al.} [MINERvA Collaboration],
    Phys.\ Rev.\ D {\bf 99}, no. 1, 012004 (2019)
    [arXiv:1811.02774 [hep-ex]].
    


          
         


       
\bibitem{Adamson:2015dkw} 
  P.~Adamson {\it et al.},
    Nucl.\ Instrum.\ Meth.\ A {\bf 806}, 279 (2016)
    [arXiv:1507.06690 [physics.acc-ph]].
    
\bibitem{Agostinelli:2002hh} 
  S.~Agostinelli {\it et al.} [GEANT4 Collaboration],
    Nucl.\ Instrum.\ Meth.\ A {\bf 506}, 250 (2003).
    
\bibitem{Allison:2006ve} 
  J.~Allison {\it et al.},
    IEEE Trans.\ Nucl.\ Sci.\  {\bf 53}, 270 (2006).
    
\bibitem{Aliaga:2016oaz} 
  L.~Aliaga {\it et al.} [MINERvA Collaboration],
    Phys.\ Rev.\ D {\bf 94}, no. 9, 092005 (2016)
  Addendum: [Phys.\ Rev.\ D {\bf 95}, no. 3, 039903 (2017)]
    [arXiv:1607.00704 [hep-ex]].
    
\bibitem{Stowell:2019zsh} 
  P.~Stowell {\it et al.} [MINERvA Collaboration],
    Phys.\ Rev.\ D {\bf 100}, no. 7, 072005 (2019)
    [arXiv:1903.01558 [hep-ex]].
    
\bibitem{Valencia:2019mkf} 
  E.~Valencia {\it et al.} [MINERvA Collaboration],
    arXiv:1906.00111 [hep-ex].
    
\bibitem{Aliaga:2013uqz} 
  L.~Aliaga {\it et al.} [MINERvA Collaboration],
    Nucl.\ Instrum.\ Meth.\ A {\bf 743}, 130 (2014)
    [arXiv:1305.5199 [physics.ins-det]].
    
\bibitem{Michael:2008bc} 
  D.~G.~Michael {\it et al.} [MINOS Collaboration],
    Nucl.\ Instrum.\ Meth.\ A {\bf 596}, 190 (2008)
    [arXiv:0805.3170 [physics.ins-det]].
      
\bibitem{Andreopoulos:2009rq} 
  C.~Andreopoulos {\it et al.},
    Nucl.\ Instrum.\ Meth.\ A {\bf 614}, 87 (2010)
    [arXiv:0905.2517 [hep-ph]].
    
\bibitem{Gran:2017psn} 
  R.~Gran,
    arXiv:1705.02932 [hep-ex].
      
\bibitem{Nieves:2011pp} 
  J.~Nieves, I.~Ruiz Simo and M.~J.~Vicente Vacas,
    Phys.\ Rev.\ C {\bf 83}, 045501 (2011)
    [arXiv:1102.2777 [heph]].
      

\bibitem{Gran:2013kda} 
  R.~Gran, J.~Nieves, F.~Sanchez and M.~J.~Vicente Vacas,
    Phys.\ Rev.\ D {\bf 88}, no. 11, 113007 (2013)
    [arXiv:1307.8105 [hep-ph]].
      
\bibitem{Schwehr:2016pvn} 
  J.~Schwehr, D.~Cherdack and R.~Gran,
    arXiv:1601.02038 [hep-ph].
    
\bibitem{Rodrigues:2015hik} 
  P.~A.~Rodrigues {\it et al.} [MINERvA Collaboration],
    Phys.\ Rev.\ Lett.\  {\bf 116}, 071802 (2016)
  Addendum: [Phys.\ Rev.\ Lett.\  {\bf 121}, no. 20, 209902 (2018)]
    [arXiv:1511.05944 [hep-ex]].
    
\bibitem{Rodrigues:2016xjj} 
  P.~Rodrigues, C.~Wilkinson and K.~McFarland,
    Eur.\ Phys.\ J.\ C {\bf 76}, no. 8, 474 (2016)
    [arXiv:1601.01888 [hep-ex]].
    
          
\bibitem{DAgostini:1994fjx} 
  G.~D'Agostini,
    Nucl.\ Instrum.\ Meth.\ A {\bf 362}, 487 (1995).
         
\bibitem{DAgostini:2010hil} 
  G.~D'Agostini,
    arXiv:1010.0632 [physics.data-an].
       
\bibitem{Adye:2011gm} 
  T.~Adye,
      arXiv:1105.1160 [physics.data-an].
      
            
           
          
  
         

         
           
          
       
\bibitem{Puckett:2011xg} 
  A.~J.~R.~Puckett {\it et al.},
    Phys.\ Rev.\ C {\bf 85}, 045203 (2012)
    [arXiv:1102.5737 [nucl-ex]].
    \bibitem{Meyer:2016oeg} 
  A.~S.~Meyer, M.~Betancourt, R.~Gran and R.~J.~Hill,
    Phys.\ Rev.\ D {\bf 93}, no. 11, 113015 (2016)
   [arXiv:1603.03048 [hep-ph]].
    \bibitem{Qattan:2015qxa} 
  I.~A.~Qattan, J.~Arrington and A.~Alsaad,
    Phys.\ Rev.\ C {\bf 91}, no. 6, 065203 (2015)
    [arXiv:1502.02872 [nucl-ex]].
    
\bibitem{Benhar:2006er} 
  O.~Benhar, D.~Day and I.~Sick,
    nucl-ex/0603032.
      
\bibitem{Benhar:2006wy} 
  O.~Benhar, D.~day and I.~Sick,
    Rev.\ Mod.\ Phys.\  {\bf 80}, 189 (2008)
  doi:10.1103/RevModPhys.80.189
  [nucl-ex/0603029].
      
\bibitem{Zeller:1973ge} 
  D.~Zeller,
    DESY-F23-73-2.
    
\bibitem{Barreau:1983ht} 
  P.~Barreau {\it et al.},
    Nucl.\ Phys.\ A {\bf 402}, 515 (1983).
  doi:10.1016/0375-9474(83)90217-8
      
\bibitem{OConnell:1987kww} 
  J.~S.~O'Connell {\it et al.},
    Phys.\ Rev.\ C {\bf 35}, 1063 (1987).
  doi:10.1103/PhysRevC.35.1063
    
\bibitem{Bagdasaryan:1988hp} 
  D.~S.~Bagdasaryan {\it et al.},
    YERPHI-1077-40-88.
      
\bibitem{Sealock:1989nx} 
  R.~M.~Sealock {\it et al.},
    Phys.\ Rev.\ Lett.\  {\bf 62}, 1350 (1989).
  doi:10.1103/PhysRevLett.62.1350
    
\end{thebibliography}
\end{document}


\thispagestyle{empty}

{\normalsize \appendix{Appendix: Supplementary Material}\hfill\vspace*{4ex}}

\begin{figure*}[h]                         \includegraphics[width=.495\linewidth]{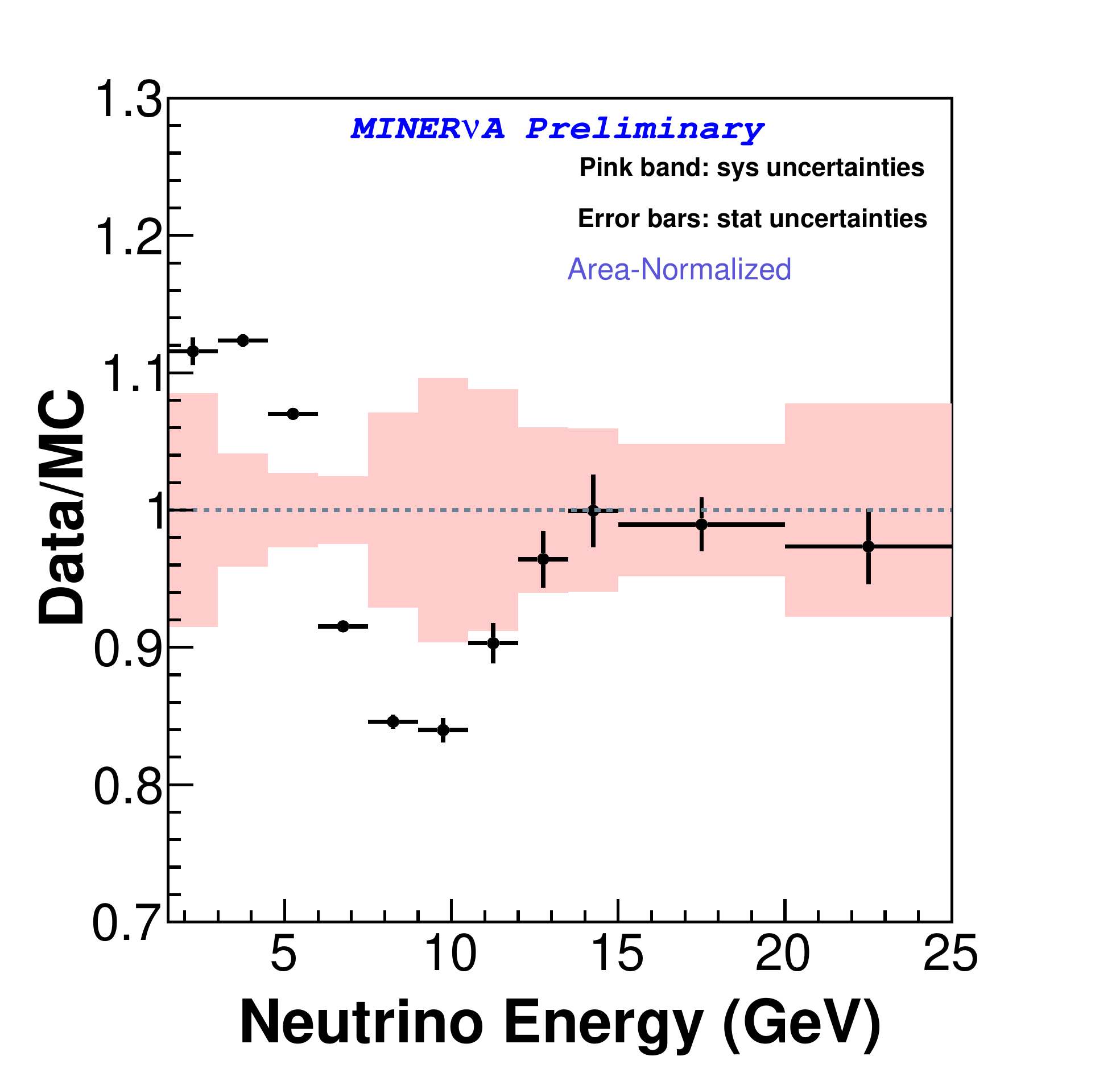}
\includegraphics[width=.495\linewidth]{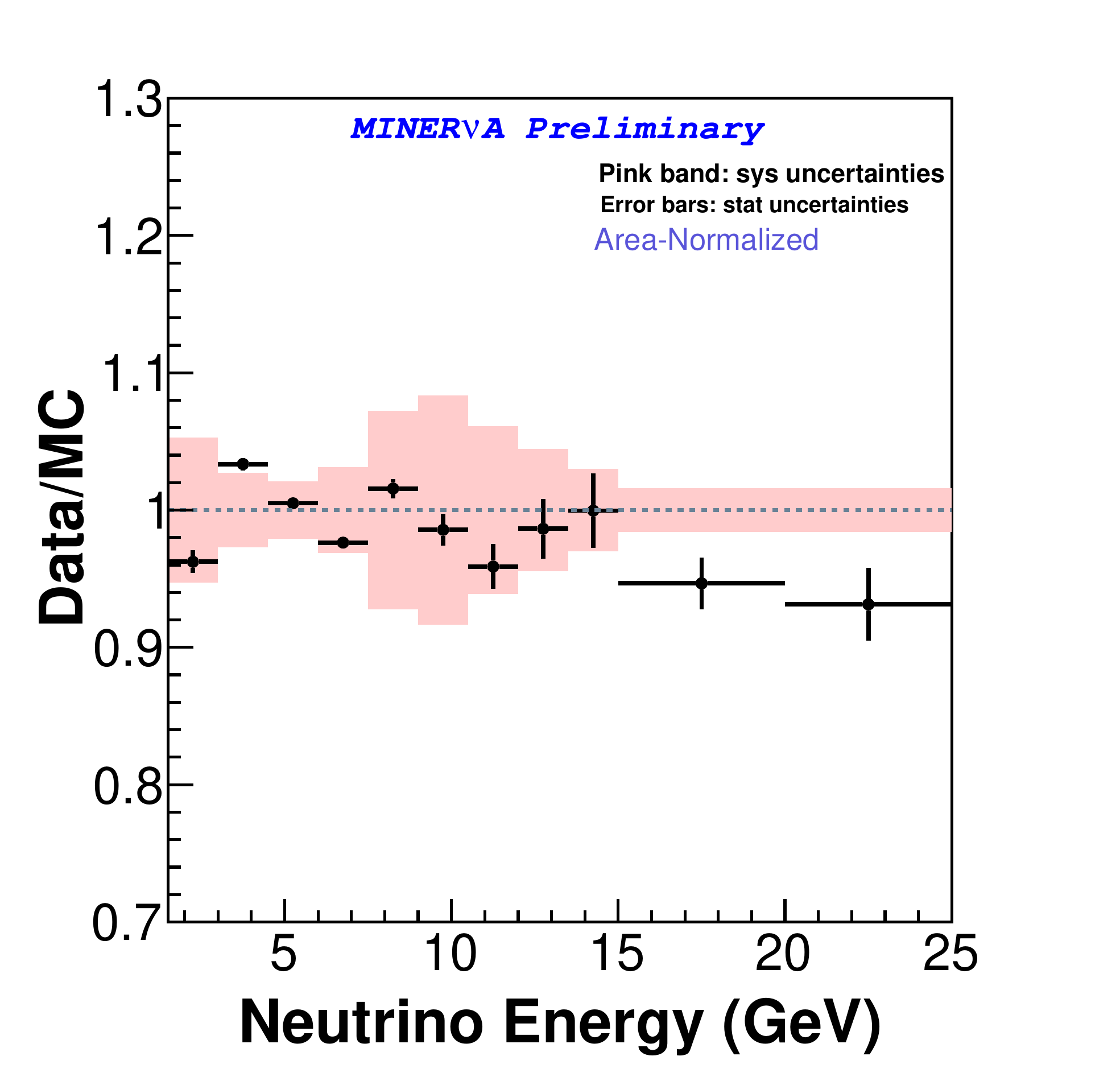}
    \caption{Ratio of data to simulation as a function of reconstructed neutrino energy for events with less than 800~MeV of hadronic energy, before (left) and after (right) a fit which allowed flux and energy scale uncertainties to vary, as described in the body of the Letter.   The simulation has been normalized to the same number of events as the data. The black error bars correspond to the statistical uncertainty on the ratio and the pink band represents the shape component of the simulation's systematic uncertainty.  The ratio is shown for the first third of the POT collected but remained consistent throughout the entire run.     }
     \label{Sys2D} 
\end{figure*}

\begin{figure*}[h]  
\includegraphics[width=.495\linewidth]{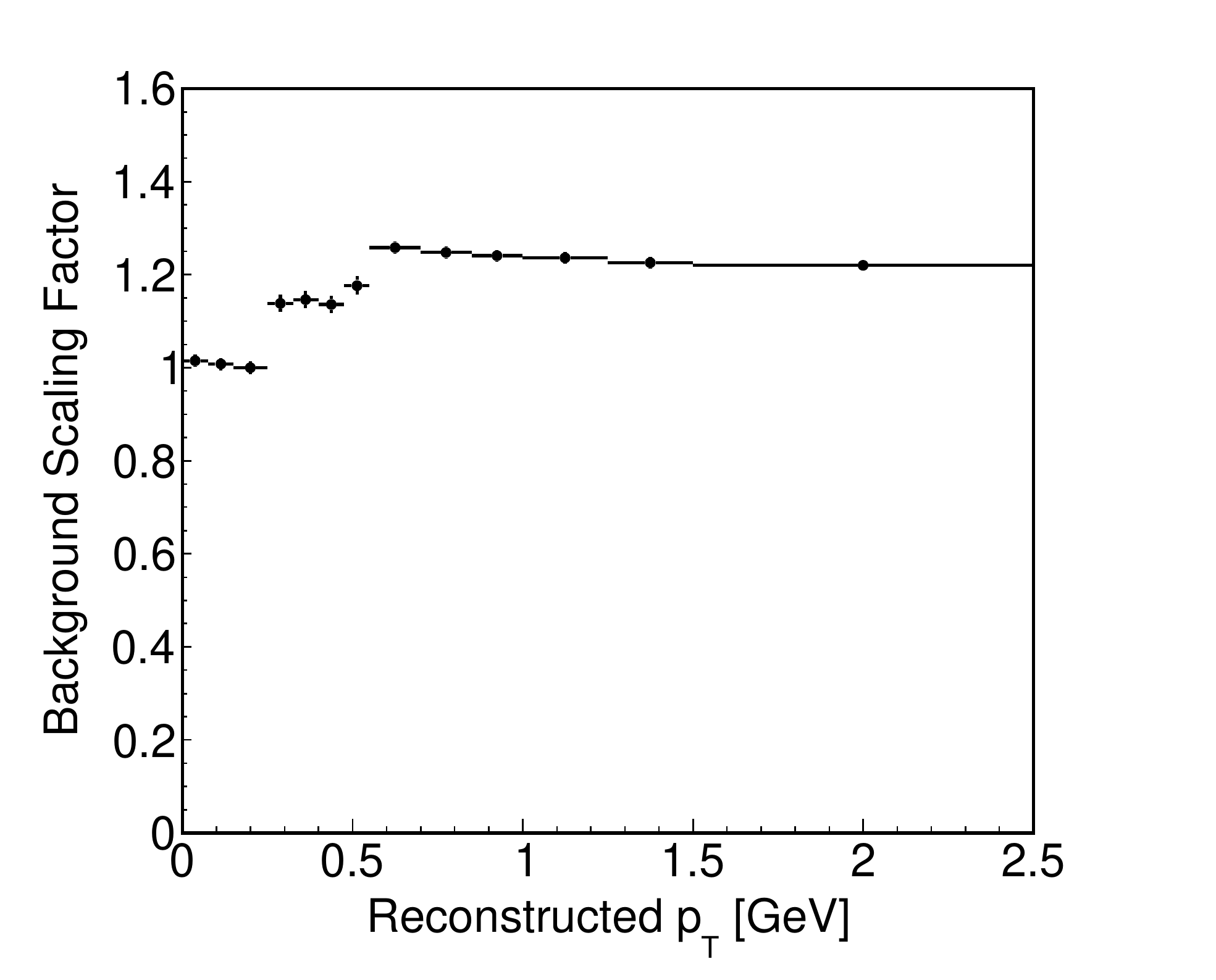}
\includegraphics[width=.495\linewidth]{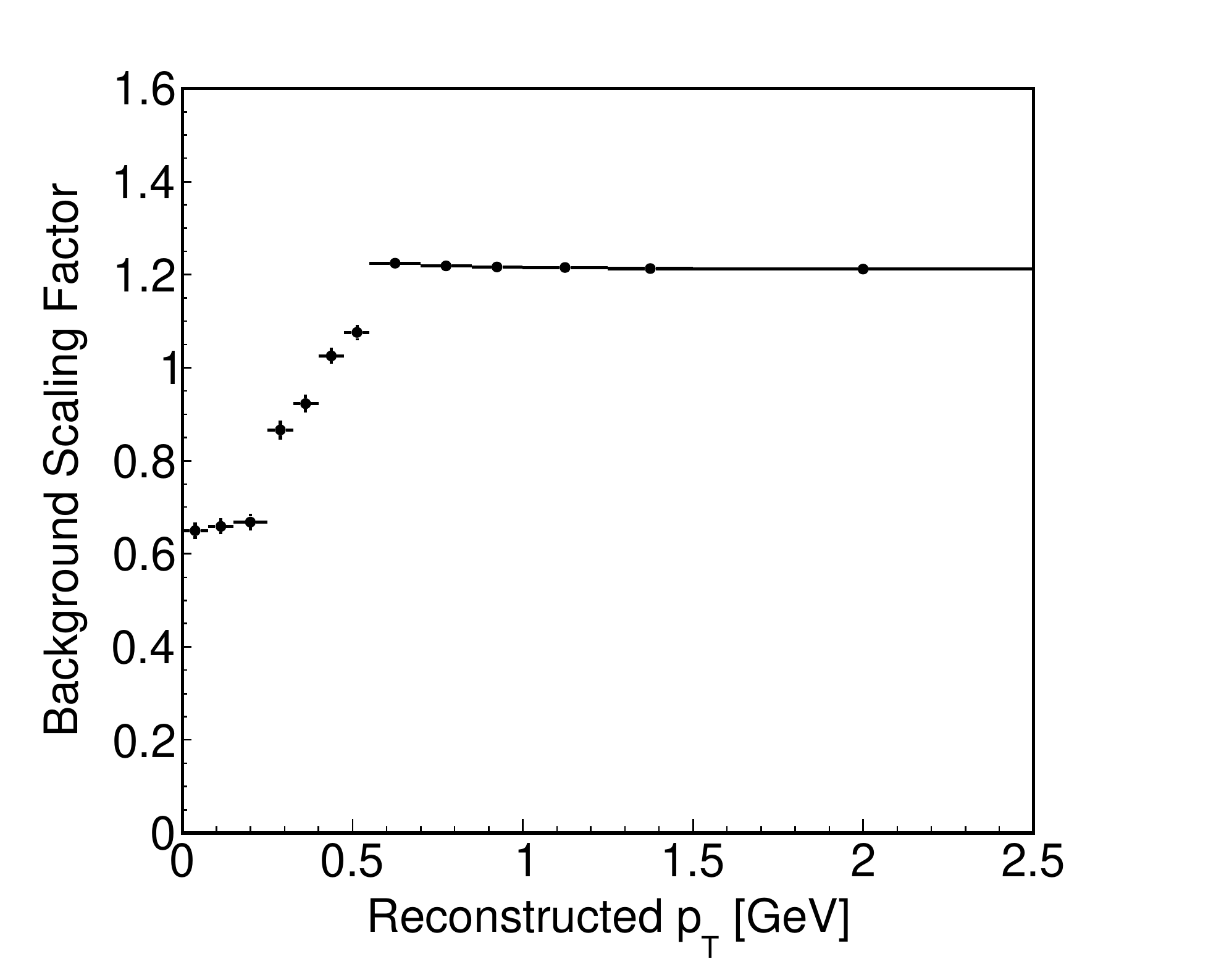}
    \caption{ Ratio of backgrounds versus reconstructed muon transverse momentum before and after the background fits described in the letter.      }
     \label{Sys2D} 
\end{figure*}

\begin{figure*}                         \includegraphics[width=\linewidth]{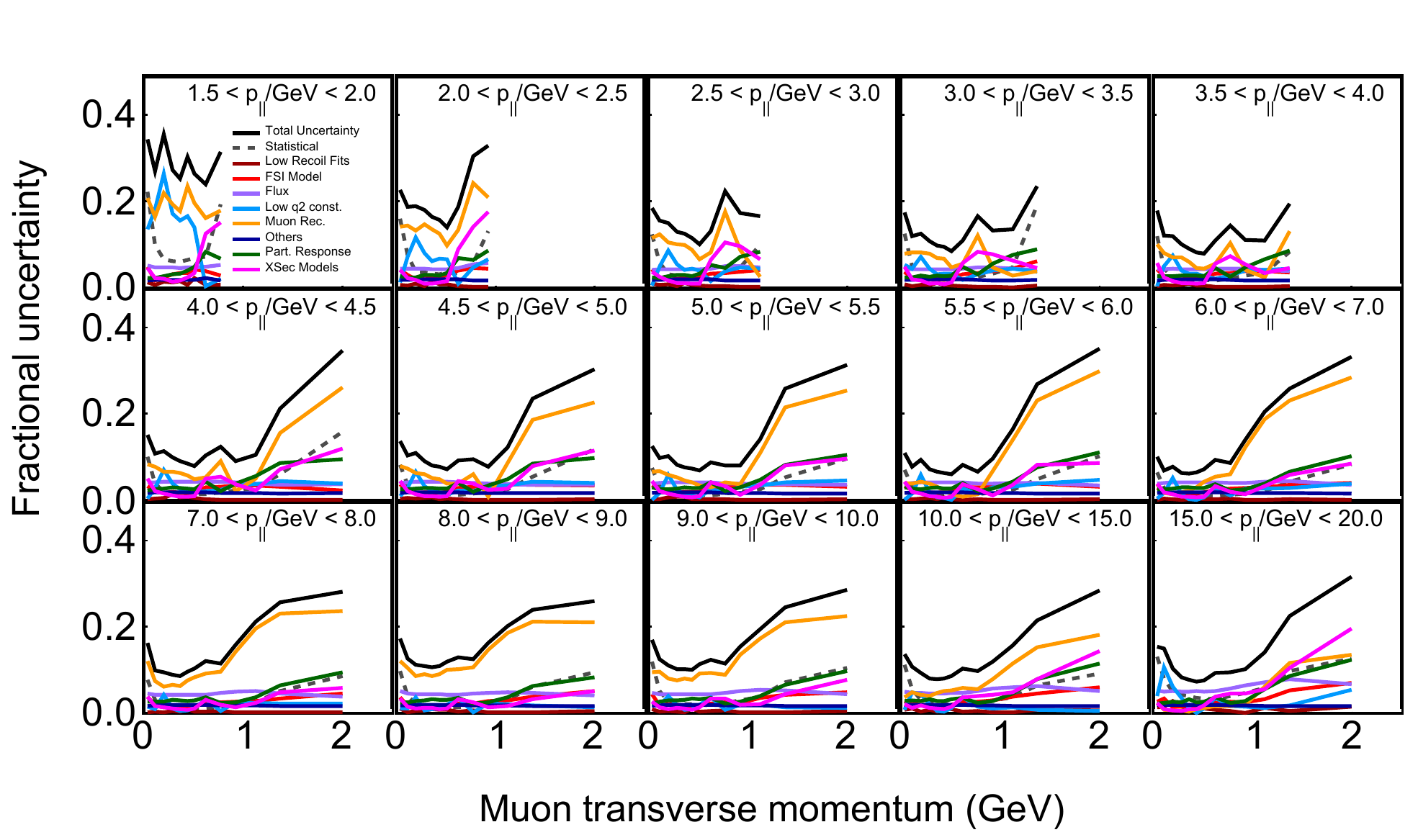}
    \caption{Fractional systematic uncertainty on the double-differential cross section as a function of $p_\perp$ and $p_\parallel$.}
     \label{Sys2D} 
\end{figure*}

\begin{figure}                         \includegraphics[trim={0cm 0cm 0cm 1.20cm},clip,width=\linewidth]{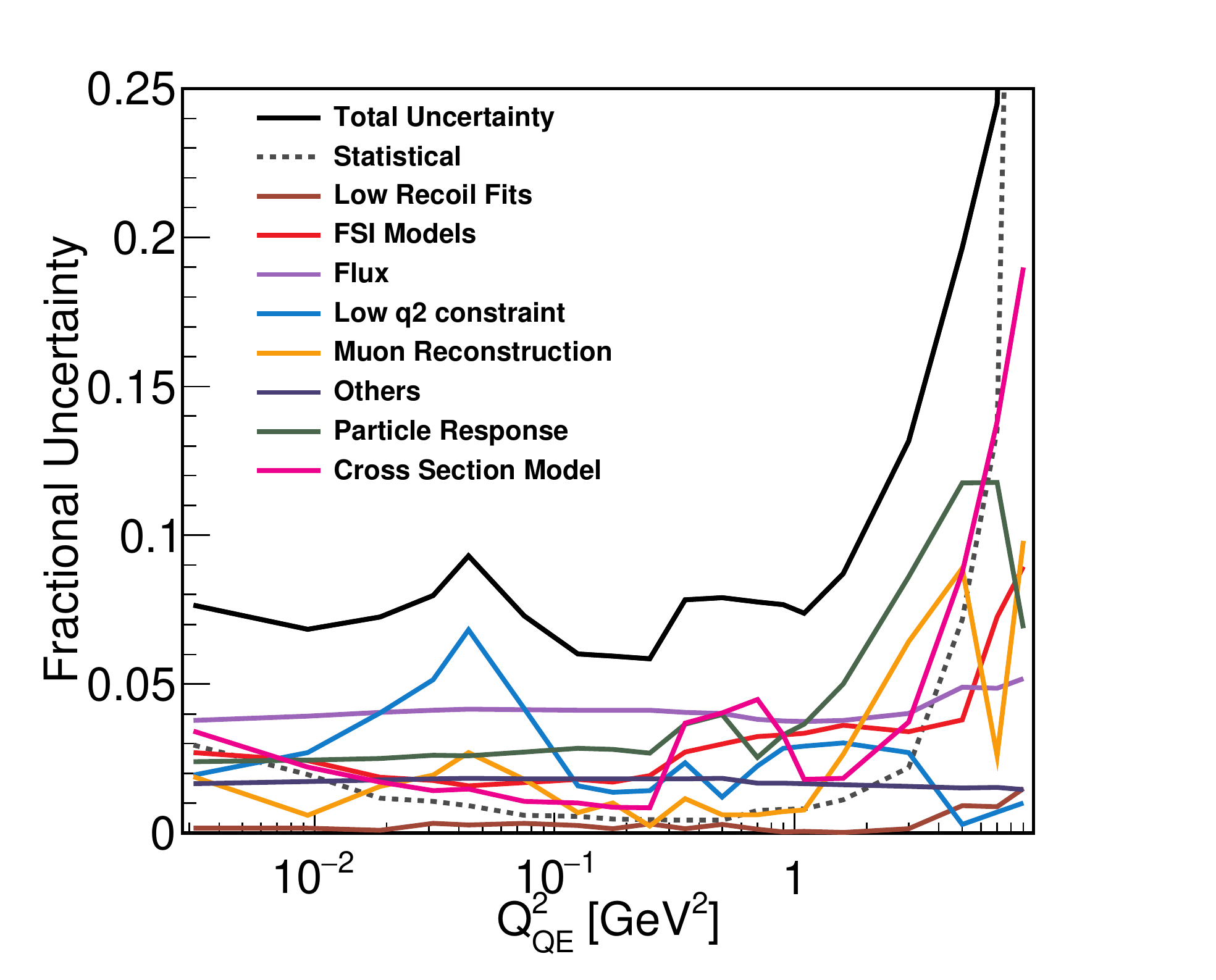}
    \caption{Fractional systematic uncertainty on the single-differential cross section as a function of $Q^2_{QE}$.}
     \label{Sys2D} 
\end{figure}

\begin{table}
\renewcommand*{\arraystretch}{1.2}
\begin{ruledtabular}
\begin{tabular}{lcc}

    &  $\chi^2$ &   $\chi^2$ \\ 
Model & Linear & Log \\
\hline
GENIE 2.12.6 &	        292 &	        372 \\
\qquad+RPA+$\pi$ tune &	        135 &	        243 \\
\qquad+RPA+$\pi$ tune+MINOS $\pi$ low $Q^2$ sup. &	        120 &	        231 \\
\hline
GENIE 2.12.6 + 2p2h &	        909 &	        640 \\
\qquad+RPA+$\pi$ tune+recoil fit (MnvGENIEv1) &	        436 &	        344 \\
\qquad+RPA+$\pi$ tune &	        442 &	        391 \\
\qquad+recoil fit+RPA+$\pi$ tune+MINOS $\pi$ low $Q^2$ sup. &	        216 &	        206 \\
\qquad+recoil fit+$\pi$ tune &	       1019 &	        719 \\
\qquad+recoil fit+RPA+$\pi$ tune+MINERvA $\pi$ low $Q^2$ sup. (MnvGENIEv2) &	        199 &	        195 \\
\hline
NuWro SF &	        498 &	        441 \\
NuWro LFG &	        363 &	        316 \\
\hline
GiBUU &	        353 &	        271 \\

\end{tabular}
\end{ruledtabular}
\caption{\label{1DChi2} $\chi^2$ of various model variants compared to the differential cross section as a function of $Q^2$. Both standard and log-normal $\chi^2$ are shown, and the number of degrees of freedom for each comparison is 19.}
\end{table}